\documentclass[a4paper,11pt]{article}

\usepackage{jheppub} 

\usepackage{yfonts}
\usepackage{amsmath}
\usepackage{amssymb}
\usepackage{mathrsfs}
\usepackage{graphicx}
\usepackage{amsfonts}
\usepackage{tikz}
\usepackage{simplewick}
\usepackage{ulem}

\newcommand{\be}{\begin{equation}}
\newcommand{\ee}{\end{equation}}
\newcommand{\bea}{\begin{eqnarray}}
\newcommand{\eea}{\end{eqnarray}}

\newcommand{\ba}{\begin{aligned}}
\newcommand{\ea}{\end{aligned}}
\newcommand{\p}{\partial}

\newcommand{\mt}[1]{\textrm{\tiny #1}}

\title{BMS$_{3}$ fermionic localization}

\author[a]{Joan Sim\'on}
\author[a,b]{and Boyang Yu}

\affiliation[a]{School of Mathematics and Maxwell Institute for Mathematical Sciences, University of Edinburgh, Edinburgh EH9 3FD, UK}
\affiliation[b]{Center for High Energy Physics, Peking University, No.5 Yiheyuan Rd, Beijing 100871, P.R. China}

\emailAdd{j.simon@ed.ac.uk}
\emailAdd{v1byu33@ed.ac.uk}

\abstract{We consider the geometric action formulation for 3d pure gravity with vanishing cosmological constant. We use fermionic localization to compute the exact torus partition function for a constant representative coadjoint orbit of $\widehat{\text{BMS}}_3$. This allows us to discuss its
1-loop exactness.  
}

\begin{document}
\maketitle
\flushbottom

\section{Introduction}

Much progress has been achieved in the understanding of quantum gravity in low spacetime dimensions due to the development of quantization methods. For example, leveraging the Chern-Simons (CS) formulation of AdS$_3$ pure gravity enables a rigorous exploration of boundary conditions, gauge fixing, and the presence of non-trivial holonomies in the CS connection \cite{Achucarro:1986uwr,Witten:1988hc}. After performing the Hamiltonian reduction, the Chern-Simons action reduces to a Wess-Zumino-Witten (WZW) boundary action \cite{Elitzur:1989nr,Coussaert:1995zp}.

Simultaneously, Alekseev and Shatashvili applied the coadjoint orbit method to study the Virasoro group \cite{Alekseev:1988ce} and reproduced the same action by deforming the geometric action associated to the coadjoint orbit with a Hamiltonian \cite{Alekseev:1990mp}.  Since the coadjoint orbits of the Virasoro group form symplectic spaces \cite{Kirillov}, these can be quantized using geometric quantization \cite{Witten:1987ty} or phase space path integrals \cite{Alekseev:1988vx}. This connection led to the
coadjoint orbit quantization of AdS$_3$ \cite{Cotler:2018zff}. In recent years, such connection was generalized to 3d Minkowski \cite{Barnich:2014kra,Barnich:2015uva,Oblak:2016eij,Barnich:2017jgw}, dS$_3$ \cite{Cotler:2019nbi}, AdS$_3$ with Comp\`ere-Song-Strominger boundary conditions \cite{Compere:2013bya,Detournay:2024gth} or Rindler boundary conditions \cite{Afshar:2024ess}. 

This quantization technology allows to compute quantum effects in these gravitational theories\footnote{Even though the classical action of Chern-Simons action coincides with the three dimensional Einstein gravity action, the quantization of both theories is different \cite{Witten:1988hc}. For example, metrics should be invertible in gravity, whereas there is no analogous requirement in the Chern-Simons gauge theory. As a result, the Chern-Simons path integral will include field configurations which cannot be interpreted as three-dimensional metrics. However, if one focuses on perturbations around a sensible (invertible) metric, the Chern-Simons theory still provides a valid description of the gravity theory. The latter is the approach followed here.}. This includes loop corrections to partition functions \cite{Maloney:2007ud,Giombi:2008vd}, correlation functions \cite{Bhatta:2016hpz,Cotler:2018zff,Merbis:2019wgk} or even entanglement entropy measures \cite{Ammon:2013hba,deBoer:2013vca,Castro:2014tta,Bagchi:2014iea,Basu:2015evh}. A further development is the \textit{1-loop exactness} of the path integrals over these geometric actions under some conditions. This was established for the Schwarzian theory \cite{Stanford:2017thb}, AdS$_3$ gravity \cite{Cotler:2018zff} and the BMS$_2$ Schwarzian theory \cite{Afshar:2021qvi}.

The purpose of this note is to explore the geometric action formulation for 3d pure Einstein gravity with vanishing cosmological constant to ask whether the one-loop contribution to the torus partition function is exact\footnote{Perturbative computations can be found in \cite{Barnich:2015mui,Garbarz:2015lua,Merbis:2019wgk,Leston:2023ugd}.}. To answer this question, we use fermionic localization.

Localization is a powerful technique to compute exact quantities in supersymmetric quantum field theories \cite{Duistermaat:1982vw,Atiyah:1984px,Witten:1988ze}, such as partition functions, Wilson loops, and other observables in several dimensions, including applications to gauge theories, string theory, and black hole entropy \cite{Pestun:2007rz,Hama:2011ea,Closset:2012ru,Benini:2015eyy,Iliesiu:2022kny}. The key point is to localize the path integral to a finite-dimensional subset of field configurations. This is achieved by deforming the action with a Q-exact term $QF$, where Q is a supersymmetry generator, that ensures the integral is dominated by the fixed points of $QF$.

This localization method has also been applied to the Schwarzian theory \cite{Stanford:2017thb} and AdS$_3$ gravity \cite{Cotler:2018zff} relying on their phase space being K$\Ddot{\text{a}}$hler. Even though it is not known to us whether such structure exists for the phase space of $\widehat{\text{BMS}}_3$, we will construct such $QF$-term using field theory techniques, enabling us to perform the path integral exactly and to probe the one-loop exactness of the torus partition function for BMS$_3$ gravity.

This paper is organized as follows. In section \ref{sec:review}, we briefly review the fermionic localization technology used in the main text. In section \ref{sec:Virasoro}, we rederive the 1-loop exactness for the Virasoro partition function using the same logic and tools that we apply later in section \ref{sec:BMS} to compute the torus BMS$_3$ partition function.  In section \ref{sec:summary}, we summarize our results. The supersymmetry of the geometric actions used to perform our calculations is discussed in appendix \ref{sec:susy-check}.

{\bf Note Added:} While finishing our work, reference \cite{Cotler:2024cia} appeared. As part of the results presented in \cite{Cotler:2024cia}, it is also claimed, and shown, that the one-loop partition function of 3d pure gravity with a vanishing cosmological constant is 1-loop exact around the Minkowski vacuum. Our results are consistent, but derived using fermionic localization rather than performing the direct path integral by observing the linear functional dependence on the superrotation variable, and applicable to other gravitational saddles, such as conical defects and flat space cosmologies.

\section{Fermionic localization}
\label{sec:review}

The technique  we use to compute the torus partition function is the method of fermionic localization. This is briefly reviewed below following \cite{Cotler:2018zff}. 

Given a symplectic phase space $\mathcal{M}$\footnote{In this note, $\mathcal{M}$ will correspond to the coadjoint orbit of the asymptotic symmetry G preserving a set of gravitational boundary conditions. In physics terminology, this corresponds to the phase space of physical configurations connected to a given classical saddle by the set of large gauge transformations preserving the asymptotic boundary conditions.}, the quantization of the classical theory leads to the partition function 
\be
\mathcal{Z}=\int_{\mathcal{M}} [dx^i]\,\text{Pf}(\omega)\,e^{-S_\mt{E}}\,,
\ee
where $\text{Pf}(\omega)$ is the Pfaffian of the symplectic form $\omega$ in $\mathcal{M}$ and $S_\mt{E}$ is the Euclidean action resulting from the Wick rotation $t\to -iy$ of the 
Alekseev-Shatashvili (AS) type action\footnote{In this note, the geometric action \eqref{eq:dg-action} will correspond to the 3d pure gravity bulk action together with a boundary term to have a well defined variational principle.} \cite{Alekseev:1990mp}
\begin{equation}
      S_\mt{geometric} = \int_\gamma \left(a + H_v\right)\,dt\,,
\label{eq:dg-action}
\end{equation}
with $\omega = \delta a$ and H$_v$ the Noether charge associated with the global symmetry generated by the flow of v, i.e. $i_v \omega = dH_v$. Explicitly,
\be
S_\mt{E}=\int dy\left(i\frac{\p x^i}{\p y}a_i+H\right).
\ee
Writing the Pfaffian term as an integral over the Grassmann-odd ghost fields $\psi^i$, the resulting path integral becomes
\be
\mathcal{Z}=\int [dx^i][d\psi^i]\,e^{-S_\mt{E}'},\qquad \text{with} \quad S_\mt{E}'\equiv S_\mt{E} + S_\omega = S_\mt{E} -\frac{1}{2}\int dy\,\omega_{ij}\psi^i\psi^j.
\label{eq:Zt1}
\ee
The total action $S_\mt{E}'$ is invariant under the symmetry generated by the  Grassmann-odd supercharge $Q$ whose actions on the dynamical fields is
\be
Qx^i=\psi^i,\quad Q\psi^i=V^i \equiv -\omega^{ij}\frac{\delta S_\mt{E}}{\delta x^i}\,.
\label{Vi}
\ee
According to the Duistermaat-Heckman theorem \cite{Duistermaat:1982vw}, the integral of a function which is both Q-exact and Q-closed vanishes. As a consequence, deforming $S_\mt{E}'\to  S_\mt{E}' + sQF$ in \eqref{eq:Zt1}, with $QF$ satisfying 
\begin{equation}
  Q^2F=0 \qquad \text{and} \qquad   (QF)_{\mt{bosonic}}\geq 0,
\label{eq:DH-cond}
\end{equation} 
does not modify the path integral, i.e.
\be
\mathcal{Z}=\mathcal{Z}[s]:=\int [dx^i][d\psi^i]\,e^{-(S'_\mt{E}+sQF)}
\label{eq:sQF}
\ee
Since the $s\to\infty$ limit localizes the path integral  to the localization manifold 
\begin{equation}
  \mathcal{M}_{\mt{loc}}=\{x_c^i\,\,| \,\, (QF)_{\mt{bosonic}}[x_c]=0,\psi^i=0\}\,,
\end{equation}
and all higher loop contributions are suppressed compared to the one-loop term, one reaches the conclusion \cite{Banerjee:2009af,Dabholkar:2010uh}
\be\label{Z-local}
\mathcal{Z}=\lim_{s\to\infty}\mathcal{Z}[s]_{\mt{1-loop}}=\int_{\mathcal M_{\mt{loc}}}[dx_c]\,e^{-S_{\mt E}[x_c]}\frac{1}{\text{SDet}'(QF)_{x_c}}
\ee
where SDet is the superdeterminant given by the ratio of the bosonic and fermionic determinants at 1-loop and the prime means the zero modes, which belong to $\mathcal{M}_{\mt{loc}}$, must be excluded.
 
In the following, we construct the fermionic localization terms for 3d pure gravity with negative or vanishing cosmological constants, i.e. for the Virasoro and BMS$_3$ groups, respectively, following a field theoretic, or cohomological, approach that does not rely on the existence of a positive definite metric on the relevant group space. This will allow us to prove the 1-loop exactness of the torus partition function for both theories.


\section{AdS$_3$ gravity}
\label{sec:Virasoro}

Let us review the application of the geometric action formulation \eqref{eq:Zt1} to 3d pure Einstein gravity with a negative cosmological constant. This technology is well known in the literature, though we believe our fermionic localization calculations in subsection \ref{sec:Ads-loc} confirming the 1-loop exactness of these partition functions are new. Below, we mainly follow \cite{Cotler:2018zff}.

All 3d pure gravity classical configurations are locally AdS$_3$ \cite{deser1984three}. Imposing Brown-Henneaux (BH) boundary conditions \cite{Brown:1986nw,Coussaert:1995zp}, the phase space of configurations is described by two functions ${\cal L}(t,\varphi),\bar {\cal L}(t,\varphi)$ \cite{Banados:1998gg,Skenderis:1999nb}
\be
ds^2=\frac{dr^2}{r^2}+(r^2+\text{G}^2{\cal L}\bar {\cal L})\,dxd\bar x+\text{G}{\cal L}\,dx^2 +\text{G}\bar {\cal L}\,d\bar x^2\,,
\ee
where $x=t+\varphi$, $\bar x=\varphi-t$, $\text{G}$ is 3d Newton's constant and $r\to\infty$ is the asymptotic boundary. 

Einstein's equations require ${\cal L}=L(x)$ and $\bar {\cal L}=\bar L(\bar x)$. The set of infinitesimal transformations preserving the BH boundary conditions is generated by \cite{Coussaert:1995zp,Brown:1986nw}
\be\label{asy-AdS}
\xi=\sigma r\,\p_r+\left(\epsilon+\frac{\bar\p\sigma}{r^2}+O(r^{-4})\right)\p+\left(\bar\epsilon+\frac{\p\sigma}{r^2}+O(r^{-4})\right)\bar\p,\qquad \sigma=-\frac{\epsilon'+\bar\epsilon'}{2}
\ee
where $\p=\p_x,\,\bar\p=\p_{\bar x}$ and $\epsilon=\epsilon(x),\,\bar\epsilon=\bar\epsilon(x)$. They close two copies of the central extension of the Virasoro algebra with equal central charges $c=\bar c=\frac{3}{2G}$ in AdS radius units \cite{Brown:1986nw}. The action of $\xi$ on the metric induces an action on the phase space $L$ and $\bar L$ given by
\be\ba
\delta L=\epsilon L'+2\epsilon' L-\frac{c}{3}\epsilon''',\quad \delta\bar L=\bar\epsilon \bar L'+2\bar\epsilon'\bar L-\frac{c}{3}\bar\epsilon'''\,.
\ea\ee
Its finite version
\be\label{AdS-finite}
\tilde L= f'^2L(f)-\frac{c}{3}\{f,x\},\quad \tilde {\bar L}= \bar f'^2\bar L(\bar f)-\frac{c}{3}\{\bar f,\bar x\}
\ee
is parameterized by two diffeomorphisms $f(x)$ and $\bar f(\bar x)$ and matches
the coadjoint action of the centrally extended Virasoro group $\widehat{\text{Vir}}$ \cite{Navarro-Salas:1999ejl,Barnich:2015uva}. Thus, the different physical configurations generated from an starting $(L_0,\,{\bar L}_0)$ belong to the same coadjoint orbit. 

There exist different inequivalent orbits \cite{Witten:1987ty}. Here, we focus on the ones labeled by constant representatives, i.e. the constant zero mode $L_0$ and ${\bar L}_0$, from the phase space functions $L(x)$ and ${\bar L}(x)$, respectively.

The relation between AdS$_3$ gravity and the technology of coadjoint orbits of $\widehat{\text{Vir}}$ can also be explicitly seen at the level of the action. Indeed, using the Chern-Simons formulation of the 3d bulk gravity theory \cite{Achucarro:1986uwr,Witten:1988hc}, including a boundary term to have a well defined variational principle, its Hamiltonian reduction consists of a sum of left-moving and right-moving parts \cite{Cotler:2018zff}. Focusing on the left-moving one, this is given by
\be
\label{action-AdS}
S_{\mt{CS}}[f,j_0]=\int dt\int^{2\pi}_0d\varphi  \left(j_0\,f'(f'+\dot f)+\frac{c}{48\pi}\frac{f''(f''+\dot f')}{f'^2}\right)
\ee
where $f'=\p_\varphi f$, $\dot f=\p_tf$ and $L_0=8\pi j_0$. This matches the Alekseev-Shatashvili action \cite{Alekseev:1990mp} and provides a particular example of \eqref{eq:dg-action}
\begin{equation}
  S_{\mt{CS}}[f,j_0] =\int_\gamma(a+H_v)dt\,.   
\label{eq:ads-geometric}
\end{equation}
Here, the geometric action $I_\mt{G}=\int_\gamma a\,dt$ matches the kinematic part of the CS action, i.e.
terms involving time derivatives, while the non-kinematic part, originating from the CS boundary term, matches the Hamiltonian H$_v$. 

The coadjoint orbit is a symplectic manifold \cite{Kirillov}. Its symplectic form equals $w=da$, where $a$ is the 1-form appearing in the geometric action $I_\mt{G}$. The Hamiltonian H$_v$ generates a conserved charge along the path $\gamma$ in the coadjoint orbit, i.e. it satisfies 
\be\label{charge}
i_v\omega=dH_v,
\ee
where $v$ is given by \eqref{asy-AdS} with $\epsilon=-1,\bar\epsilon=0$. 

Once a specific subspace of configurations of the full 3d gravity is identified with the coadjoint orbit of $L_0$\footnote{To properly account for AdS$_3$, one must add the contribution from the right sector labeled by ${\bar L}_0$.}, the path integral techniques reviewed in section \ref{sec:review} can be applied. Concretely, given the relation between the geometric and the CS action \eqref{eq:ads-geometric}, the one-form $a$ is given by the kinematic part with $\dot f$ replaced by $\delta f$,
\be
a=\int_0^{2\pi}d\varphi \left(j_0\,f'\delta f+\frac{c}{48\pi}\frac{f''\delta f'}{f'^2}\right).
\ee
Consequently, the symplectic form $\omega$ is computed to be
\be\label{sym-Vir}
\omega=\int^{2\pi}_0d\varphi\,\left( j_0\,\delta f'\wedge\delta f+\frac{c}{48\pi}\frac{\delta f''}{f'^2}\wedge\delta f'\right)\,.
\ee
This symplectic form determines the full geometric action in \eqref{eq:Zt1} (prior to Wick rotation). This provides the starting point for our computations.

\subsection{Torus partition function at one-loop}

Before discussing localization, let us compute the one-loop torus partition function including the contribution from the Pfaffian computed explicitly. The same computation involving only the bosonic contribution can be found in \cite{Cotler:2018zff}. 

First, perform a Wick rotation $t\to-iy$ leading to the Euclidean action
\be
S_\mt{E}=\int dyd\varphi \left(j_0\,f'(f'+i\p_yf)+\frac{c}{48\pi}\,{f''(f''+i\p_yf')}{f'^2}\right)\,.
\label{eq:Vir-E}
\ee
The ghost action $S_\omega$ in \eqref{eq:Zt1} is obtained from the symplectic form \eqref{sym-Vir} and is given by
\be
S_\omega=\int dyd\varphi \left( j_0\,\psi\psi'+ \frac{c}{48\pi}\,\frac{\psi'\psi''}{f'^2}\right)\,.
\label{eq:Vir-w}
\ee

Given a torus with cycles $(\varphi,y)\sim(\varphi+2\pi,y)\sim(\varphi+\beta\Omega,y+\beta)$, the phase space functions $f(\varphi,y)$ and $\psi(\varphi,y)$ satisfy the boundary conditions
\be\ba
&f(\varphi+2\pi,y)=f(\varphi,y)+2\pi,\quad f(\varphi+\beta\Omega,y+\beta)=f(\varphi,y)
\\&\psi(\varphi+2\pi,y)=\psi(\varphi,y),\quad \psi(\varphi+\beta\Omega,y+\beta)=\psi(\varphi,y).
\ea
\ee
The torus partition function is a path integral over the phase space given by \eqref{eq:Zt1} 
\be
\mathcal{Z}=\int [Df] [D\psi]\, e^{-(S_\mt{E}+S_\omega)}.
\ee
The saddle solution to the action $S_\mt{E} + S_\omega$ is given by $f_0=\varphi-\Omega y$ and $\psi=0$ \cite{Cotler:2018zff}. The expansion of $f$ and $\psi$ into Fourier modes around this saddle is given by 
\be
\ba\label{expand}
f&=f_0+\epsilon(\varphi,y)=f_0+\sum_{m,n}\frac{\epsilon_{mn}}{(2\pi)^2}e^{-inf_0-\frac{2\pi imy}{\beta}}\\
\psi&=\sum_{m,n}\frac{\psi_{mn}}{(2\pi)^2\sqrt{\beta}}e^{-inf_0-\frac{2\pi imy}{\beta}}\,.
\ea
\ee
Due to the reality of fields $f$ and $\psi$, the real and imaginary components of these modes
\be
\epsilon_{mn}=\epsilon^\mt{R}_{mn}+i\epsilon^\mt{I}_{mn},\quad \psi_{mn}=\psi^\mt{R}_{mn}+i\psi^\mt{I}_{mn}
\label{eq:comp-conj}
\ee
satisfy $\epsilon^\mt{R}_{mn}=\epsilon^\mt{R}_{-m,-n}$ and $\epsilon^\mt{I}_{mn}=-\epsilon^\mt{I}_{-m,-n}$, with analogous conditions for the $\psi_{mn}$ modes. Hence, by defining $\epsilon_{mn}^*=\epsilon_{-m,-n}$ and $\psi^*_{mn}=\psi_{-m,-n}$, this star operation will match complex conjugation. Plugging \eqref{expand} into the action and expanding to the quadratic order, we get
\be
\ba\label{AdS:act-exp}
  S_\mt{E} &=-4\pi^2i\tau j_0+\frac{ic}{96\pi^3}\sum_{n,m}n(n^2+\frac{48\pi}{c}\, j_0)(m-n\tau)|\epsilon_{mn}|^2 \\
  S_\omega &=\frac{ic}{384\pi^4}\sum_{n,m}n(n^2+\frac{48\pi}{c}\,j_0)\psi_{mn}\wedge\psi_{mn}^*
\ea
\ee
where $\tau=\frac{\beta\Omega+i\beta}{2\pi}$ and $\epsilon_{mn}^*=\epsilon_{-m,-n},\psi^*_{mn}=\psi_{-m,-n}$, as discussed below \eqref{eq:comp-conj}. The $\epsilon$-independent piece in the action defines the saddle contribution
\begin{equation}
  S_0= S_\mt{E}(f_0) =-4\pi^2i\tau\, j_0
\label{eq:Vir-saddle}
\end{equation}
Note the Hamiltonian $\int dy\,H$ is given by the real part of $S_\mt E$ and equals
\be
\int^\beta_0 dy\,H=\frac{\beta c}{192\pi^4}\sum_{n,m}n^2(n^2+\frac{48\pi}{c}\,j_0)|\epsilon_{mn}|^2.
\ee
Convergence of the partition function requires the latter to be bounded from below, a condition that holds if and only if $j_0\geq-\frac{c}{48\pi}$. We will only consider such situation in the following. Furthermore, the summation in \eqref{expand} must exclude the modes associated with the isometry of the state. For the vacuum state with $j_0=-\frac{c}{48\pi}$, the isometry group is $\mathrm{SL}(2,\mathbb R)$ whereas for states with $j_0>-\frac{c}{48\pi}$, the isometry group is $\mathrm{U}(1)$ \cite{Witten:1987ty}. As a result, the summation in \eqref{expand} excludes $n=0,\pm1$ when $j_0=-\frac{c}{48\pi}$, and $n=0$ when $j_0>-\frac{c}{48\pi}$.

The one-loop contribution to the partition function can now be extracted from the coefficients of the quadratic terms in \eqref{AdS:act-exp}. Notice, in particular, how the contribution from the $n(n^2+\frac{48\pi j_0}{c})$ factor cancels out. This leads to the final result
\be\label{Z-Vir}
\mathcal{Z}_\mt{1-loop}=e^{-S_0}\prod_{n,m}|m-n\tau|^{-1/2}=e^{-S_0}\det(\bar\p)^{-1/2},\quad \bar\p=\p_\varphi+i\p_y\,.
\ee
After zeta-regularization, \eqref{Z-Vir} is computed to be \cite{Maloney:2007ud,Giombi:2008vd,Cotler:2018zff}
\be
\mathcal{Z}_\mt{1-loop}=q^{2\pi j_0}\prod\frac{1}{1-q^n},\quad q=e^{2\pi i\tau},
\ee
matching the holomorphic Virasoro character.

\subsection{Localization}
\label{sec:Ads-loc}

Here, we reproduce the one-loop exactness of the torus partition function using the localization arguments reviewed in section \ref{sec:review}. This requires us to discuss the supersymmetry of the action \eqref{eq:Zt1} and the construction of a localization term.

When performing the same expansion as in \eqref{expand} for the full action, the resulting action
\be
 S'_\mt{E}=S_0+\int dyd\varphi \left(j_0\,\epsilon'(i\p_y\epsilon+\epsilon')+\frac{c}{48\pi}\frac{\epsilon''(\epsilon''+i\p_y\epsilon')}{(1+\epsilon')^2}+j_0\,\psi\psi'+\frac{c}{48\pi}\frac{\psi'\psi''}{(1+\epsilon')^2}\right)
\ee
is invariant under the supersymmetry transformations
\be
Q\epsilon=\psi,\quad Q\psi=-\epsilon'-i\p_y\epsilon\,.
\ee
This is shown in appendix \ref{sec:vir-check}.

The remaining task to apply fermionic localization is to write a proper localization term. Consider the family of Q-exact terms
\be
QF=\int Q(\psi D\epsilon)
\label{eq:Vir-loc-ansatz}
\ee
Notice these are also Q-closed for any arbitrary differential operator $D$. When restricting to first order operators, i.e. $D=a_1\p_y+a_2\p_\varphi$, the localization term \eqref{eq:Vir-loc-ansatz} becomes
\be\ba
QF=&-\int dyd\varphi\, (\epsilon'+i\p_y\epsilon)(a_1\p_y\epsilon+a_2\epsilon')+\psi(a_1\p_y\psi+a_2\psi')\\
=-&\sum_{n,m}\frac{i(m-n\tau)(a_2n\beta+a_1(2\pi m+in\beta-2\pi n\tau))}{2\pi^2\beta}|\epsilon_{mn}|^2
\\&+i\frac{a_2n\beta+a_1(2\pi m+in\beta-2\pi n\tau)}{8\pi^3\beta}\psi_{mn}\wedge\psi_{mn}^*.
\ea
\ee
Positivity of its bosonic part, as in \eqref{eq:DH-cond} can be achieved by the choice $a_1=i,a_2=-1$, i.e. $D=i\p_y-\p_\varphi\equiv-\p$, leading to
\be\label{QF-Vir-final}
  QF=\int \bar\p\epsilon\p\epsilon+\psi\p\psi \qquad \Rightarrow \quad (QF)_\mt{bosonic}=\sum_{n,m}\frac{|m-n\tau|^2}{\pi\beta}|\epsilon_{mn}|^2\,.
\ee
It follows $\text{SDet}'(QF)=\det(\bar\p)^{-1/2}$. Hence, according to \eqref{Z-local}, the full partition function matches the one-loop partition function \eqref{Z-Vir}. This reproduces the one-loop exactness for the partition function of 3d gravity with negative cosmological constant around an specific saddle, i.e. constant coadjoint orbit representative \cite{Cotler:2018zff}.


\section{BMS$_3$ gravity} 
\label{sec:BMS}

In this section, we apply the same technology and logic to 3d pure Einstein gravity with vanishing cosmological constant aiming at exploring the 1-loop exactness of its torus partition function.

Asymptotically Minkowski metrics in 3d pure gravity \cite{Barnich:2010eb}
\begin{equation}
  ds^2 = {\cal M}(u,\varphi)\,du^2 -2 dr\,du + 2 {\cal N}(u,\varphi)\,dud\varphi + r^2 d\varphi^2\,,
\label{eq:flat-metric}
\end{equation}
are parameterised by two functions ${\cal M}(u,\varphi),\,{\cal N}(u,\varphi)$, with future null infinity $\mathscr{I}^+$ reached by $r\to \infty$. Einstein's equations require ${\cal M}={\cal M}(\varphi)$ and ${\cal N}={\cal L}(\varphi) + \frac{u}{2}{\cal M}^\prime(\varphi)$. Imposing boundary conditions \cite{Ashtekar:1996cd,Barnich:2006av}
\begin{equation}
\label{eq:asym-bms}
	ds^2 = O(1)\, du^2 - 2 \left(1 + O(1/r)\right) dr\,du  + O(1)\, du d\varphi + r^2 d\varphi^2\,,
\end{equation}
the set of infinitesimal transformations preserving the near null infinity behaviour of the metric is generated by the vector fields
\begin{equation}
\ba
	\xi &= ( \epsilon_\mt{L}(\varphi)+u  \epsilon_\mt{L}'(\varphi))\partial_u + \left(  \epsilon_\mt{L}(\varphi) - \frac{1}{r}(\epsilon_\mt{R}'(\varphi) + u \epsilon_\mt{L}''(\varphi))\right)\partial_{\varphi} \\
	&+\left( - r \epsilon_\mt{L}'(\varphi) +  \epsilon_\mt{R}''(\varphi) + u  \epsilon_\mt{L}'''(\varphi)\right)\partial_r \,,
\ea
\label{eq:bms-alg}
\end{equation}
up to subleading terms at large $r$. These belong to $\widehat{\mathfrak{bms}}_3$ and generate $\widehat{\text{BMS}}_3$, the central extension of BMS$_3$.

The action of $\xi$ on \eqref{eq:flat-metric} induces an action on the phase space functions ${\cal M}$ and ${\cal N}$ given by \cite{Merbis:2019wgk}
\begin{equation}
\ba
  \delta {\cal M} &= \epsilon_\mt{L} {\cal M}^\prime + 2\epsilon^\prime_\mt{L} {\cal M} - 2\epsilon^{'''}_\mt{L}\,, \\
  \delta {\cal N} &= \frac{1}{2}\epsilon_\mt{R} {\cal M}^\prime + \epsilon^\prime_\mt{R} {\cal M} + \epsilon_\mt{L} {\cal N}^\prime + 2\epsilon^\prime_\mt{L} {\cal N} - \epsilon^{'''}_\mt{R}  \,.
\ea
\label{eq:bms-coadjoint-traf}
\end{equation}
These match the infinitesimal form of the coadjoint action \cite{Barnich:2015mui,Merbis:2019wgk}. Its finite version
\be\ba\label{orbit-BMS}
\tilde{\mathcal{M}}&=f'^2\mathcal{M}(f)-2\{f,\varphi\}\\
\tilde {\mathcal{N}}&=f'^2(\mathcal N(f)+\frac{1}{2}\alpha(f)\p_f \mathcal{M}(f)+\mathcal{M}(f)\p_f\alpha(f)-\p_f^3\alpha(f)).
\ea
\ee
consists of a \textit{superrotation} $\varphi \to f(\varphi)$ on the circle, together with a \textit{supertranslation} $u \to u + \alpha(\varphi)$.

The relation between the gravitational and geometric actions reviewed for AdS$_3$ extends to this case. Indeed, the Hamiltonian reduction of the CS formulation for this theory\footnote{The first step in this reduction involving the rewriting in terms of a WZW boundary model in the specific context of 3d pure flat gravity was performed in \cite{Barnich:2013yka}.} equals \cite{Merbis:2019wgk} 
\begin{equation}
\ba 
    S_{\mt{CS}}[f,\alpha,L_0,M_0] &=  -\frac{k}{2\pi} \int d u d\varphi\, \Big[\left( L_0 + M_0 \partial_f\alpha(f) - \partial_f^3\alpha(f) \right) \dot{f} f' \Big. \\
    & \Big. -  \frac12 \left( M_0 f'^2  - 2 \{f,\varphi\} \right) \Big] \\
    &=\int_\gamma \left(a + H_v\right)\,du\,.
\ea
\label{Sfinal}
\end{equation}
where $\dot f=\p_uf$. The last equality describes the modified geometric action defined on a path $\gamma$ in the coadjoint orbit of $\widehat{\text{BMS}}_3$ labeled by constant representatives $(M_0,\,L_0)$. These are the zero modes of the phase space functions ${\cal M}(\varphi)$ and ${\cal L}(\varphi)$, respectively \cite{Barnich:2017jgw}. The Hamiltonian still satisfies \eqref{charge} with $v$ now given by \eqref{eq:bms-alg} with $\epsilon_\mt{L}=-1,\epsilon_\mt{R}=0$.

Given the above relation, one can read off the one-form $a$ to be \cite{Barnich:2017jgw}
\be
\ba
a=-\frac{k}{2\pi}\int^{2\pi}_0d\varphi\, f'(L_0+M_0\p_f\alpha-\p_f^3\alpha)\,\delta f
\ea
\label{eq:a-bms3}
\ee
Using the chain rule $\frac{d}{df}=\frac{1}{f'} \,\frac{d}{d\varphi}$, \eqref{eq:a-bms3} can be written as
\be
a=-\frac{k}{2\pi}\int_0^{2\pi} d\varphi \left( L_0f'\delta{f}+M_0{\delta f}\tilde\alpha'-\frac{\tilde\alpha'(f'\delta f''-f''\delta f')}{f'^3}\right)
\ee
where $\tilde\alpha=\alpha\circ f$. The symplectic form $\omega$ in the coadjoint orbit can now be computed by $\omega = \delta a$ leading to
\begin{equation}
\ba
  \omega = -\frac{k}{2\pi}\int_{0}^{2\pi}d\varphi &\left(L_0\,\delta f'\wedge\delta f+M_0\,\delta\tilde\alpha'\wedge\delta f  -\frac{1}{ f'}\delta\tilde\alpha'\wedge\left(\frac{\delta f'}{f'}\right)'+\tilde\alpha'\frac{\delta f'\wedge\delta f''}{f'^3}\right).
\ea
\label{eq:omega-bms3}
\end{equation}
This symplectic form determines the full geometric action in \eqref{eq:Zt1} (prior to Wick rotation). This provides the starting point for our computations.

\subsection{One-loop torus partition function}
\label{sec:1-loop}

To check whether the one-loop torus partition function of BMS$_3$ is exact, we first perform a perturbative calculation. The latter is already available in the literature, see \cite{Barnich:2015mui,Merbis:2019wgk}. 

The torus is still defined by $(\varphi,y)\sim(\varphi+2\pi,y)\sim(\varphi+\beta\Omega,y+\beta)$.
The bosonic phase space parameterised by the functions $f$ and $\alpha$ satisfies the boundary conditions
\be\ba
&f(\varphi+\Omega\beta,y+\beta)=f(\varphi,y),\quad \tilde\alpha(\varphi+\beta\Omega,y+\beta)=\tilde\alpha(\varphi,y)
\\&f(\varphi+2\pi,y)=f(\varphi,y)+2\pi,\quad \tilde\alpha(\varphi+2\pi,y)=\tilde\alpha(\varphi,y),
\ea\ee
where $\tilde\alpha(\varphi,y)\equiv \alpha\circ f(\varphi,y)= \alpha(f(\varphi,y),y)$. 

After performing the Wick rotation $u\to -iy$,
the euclidean action becomes
\be
S_\mt{E}=-\frac{k}{2\pi}\int dyd\varphi\, \left(i(L_0f'+M_0\tilde\alpha')\p_yf-i\frac{\tilde\alpha'(f'\p_yf''-f''\p_yf')}{f'^3}-\frac{M_0}{2}f'^2+\{f,\varphi\}\right)
\ee
where $k=\frac{d}{12}$. Using \eqref{eq:omega-bms3}, the ghost action reduces to
\be
S_\omega=\frac{k}{2\pi}\int dyd\varphi \left(L_0\psi'_f\psi_f+M_0\psi_\alpha'\psi_f-\frac{\psi_\alpha'(\psi_f''f'-\psi_f'f'')}{f'^3}+\frac{\tilde\alpha'\psi_f'\psi_f''}{f'^3}\right)\,,
\ee
with both ghost fields $\psi_f$ and $\psi_\alpha$ being periodic along the torus cycles. The torus partition function \eqref{eq:Zt1} can then be written as
\be
\mathcal{Z}=\int [Df][D\tilde\alpha][D\psi_f][D\psi_\alpha]\,e^{-(S_{\mt E}+S_\omega)}.
\ee
Since the BMS$_3$ Hamiltonian, which is given by the second line of \eqref{Sfinal}, equals the AdS$_3$ one upon the identification $M_0=\frac{48\pi}{c}\, j_0$, it follows the BMS$_3$ Hamiltonian is bounded from below for $M_0\geq-1$. This condition includes the Minkowski vacuum $M_0=-1$, conical deficit solutions $-1<M_0<0$ and flat space cosmologies $M_0>0$. When imposing regularity conditions on the cosmological horizon (or trivial holonomy condition in the CS formulation), $(\beta,\Omega)$ are related to $M_0,L_0$ by \cite{Bagchi:2013lma,Merbis:2019wgk}
\be
\Omega=\frac{iM_0}{L_0},\quad \beta=\frac{2\pi L_0}{M_0^{3/2}}.
\ee
The perturbative computation of the one-loop partition function depends on the value of the chemical potential
\be
\theta\equiv\frac{\beta\Omega}{2\pi}=\frac{i}{\sqrt{M_0}}.
\label{eq:theta}
\ee
This is purely imaginary for positive $M_0$, and real for $-1\leq M_0<0$. We discuss these different cases next.

\paragraph{Irrational or purely imaginary of $\theta$.} When $\theta$ is irrational or purely imaginary, there exists a unique solution to the saddle point equations compatible with periodicity
\begin{equation}
  f(\varphi,y) = f_0(\varphi,y) = \varphi - \Omega\,y\,, \qquad \alpha = \psi_f = \psi_\alpha = 0\,. 
\end{equation}
To compute the spectrum of quadratic fluctuations, we expand the fields in Fourier modes
\be
\ba\label{Fourier}
&f(\varphi,y)=f_0+\sum_{m,n}\frac{\epsilon_{mn}}{(2\pi)^2}e^{-\frac{2\pi i m y}{\beta}}e^{- inf_0},\quad \tilde{\alpha}(\varphi,y)=\sum_{m,n}\frac{\alpha_{mn}}{(2\pi)^2}e^{-inf_0}e^{-\frac{2\pi imy}{\beta}}
\\&\psi_f=\sum_{m,n}\frac{a_{mn}}{(2\pi)^2\sqrt{\beta}}e^{-inf_0}e^{-\frac{2\pi imy}{\beta}},\quad \psi_\alpha=\sum_{m,n}\frac{b_{mn}}{(2\pi)^2\sqrt{\beta}}e^{-inf_0}e^{-\frac{2\pi imy}{\beta}}.
\ea
\ee
Due to \eqref{eq:theta}, $\theta$ is irrational or purely imaginary only for non-vacuum states ($M_0>-1$). The isometry group of these states is $\mathrm{U}(1)\times R$. Since the latter should be modded out, the summation \eqref{Fourier} excludes modes with $n=0$.

The reality of fields $f$ and $\tilde\alpha$ imposes the same constraints on $\epsilon_{mn}$ and $\alpha_{mn}$, as the ones discussed below \eqref{eq:comp-conj}. Hence, we shall adopt the same definition here : $\epsilon^*_{mn}=\epsilon_{-m-n}$ and $\alpha^*_{mn} = \alpha_{-m-n}$. It follows, the action $S_\mt{E}'$ at quadratic order becomes
\be
\begin{aligned}
  S_\mt{E} &=\frac{d}{24}\beta(M_0+2i\Omega L_0) \\
   &-\frac{ik}{(2\pi)^3}\sum_{m,n}\left[(L_0n(m-n\theta)+\frac{i\beta}{4\pi}n^2(n^2+M_0))|\epsilon_{mn}|^2+(m-\theta n)(n^3+M_0n)\epsilon^*_{mn}\alpha_{mn}\right],
\label{SE-bo}
\end{aligned}
\ee
where the first line defines the value of the Euclidean action at the saddle point $f_0$ 
\begin{equation}
    S_0=S_\mt{E}(f_0) = \frac{d}{24}\beta(M_0+2i\Omega L_0)
\end{equation} 
and 
\be\label{Somega}
  S_\omega=\frac{ik}{(2\pi)^4}\left(n L_0\, a_{mn}\wedge a^*_{mn} + n(M_0+n^2)\,b_{mn}\wedge a^*_{mn}\right)\,.
\ee
Before computing the 1-loop determinant, we comment on dimensions. Since $y\sim L$ (for some length scale $L$), $k\sim L^{-1}$ and $\varphi$ is dimensionless, i.e. $\varphi \sim L^0$, it follows $\beta\sim L$ and $\Omega\sim L^{-1}$. Since the action is dimensionless,  $M_0\,,\epsilon_{mn}\,,a_{mn}\sim L^0$ are
dimensionless, while $L_0\,,\alpha_{mn}\,,b_{mn}\sim L$. Finally, since the partition function should also be dimensionless, the measure in the path integral, up to dimensionless numerical factors, should be
\be
[d\epsilon][d\tilde\alpha][d\psi_f][d\psi_\alpha]=\prod_{mn}d\epsilon_{mn}{d\alpha_{mn}}d\tilde a_{mn}d\tilde b_{mn},
\ee
with $d\tilde\alpha_{mn}=L^{-1}\,d\alpha_{mn}$ and $d\tilde b_{mn}=L\,db_{mn}$. Note that since $\psi_\alpha$ is a ghost field, $d\psi_\alpha$ has the opposite dimension to $\psi_\alpha$. Since we shall not be specific about numerical factors, we choose $L=k^{-1}$.

The 1-loop partition function is obtained by evaluating the Gaussian functional integrals in \eqref{SE-bo} and \eqref{Somega}. Notice how the contributions from $n(n^2+M_0)$ cancel, leading to the result
\be\label{Z-BMS}
\mathcal{Z}_{\mt{1-loop}}=e^{-S_0}\prod_{m,n}(m-n\theta)^{-1}.
\ee
After zeta-regularization, the one-loop partition function agrees with the BMS$_3$ character in the induced representation \cite{Barnich:2015mui,Merbis:2019wgk}
\be
\mathcal{Z}_{\mt{1-loop}}=e^{-S_0}\prod_n\frac{1}{|1-q^n|^2},\qquad q=e^{2\pi i\theta}.
\ee

\paragraph{Rational values of $\theta$.}  The computation of the one-loop partition function is more involved for two reasons. First, there is no unique saddle point. For example, there exists a family of saddles given by
\be\label{new-saddle}
   f(\varphi,y) = f_0(\varphi,y) = \varphi - \Omega\,y\,,\qquad\tilde\alpha(\varphi)=\tilde\alpha(\varphi+2\pi)=\tilde\alpha(\varphi+2\pi\theta)\,.
\ee
However, the full characterization of saddles requires to solve nonlinear ODEs obtained by varying $S_{\mt E}$ with respect to $f$ and $\tilde\alpha$, together with imposing the appropriate boundary conditions. Second, when evaluating the contribution to these saddle points, there can exist zero modes making the Hessian of $S_{\mt E}$ degenerate. These require careful treatment. 

It is still instructive to compute the contribution from the saddle $f=f_0,\,\tilde\alpha=0$, as done in the irrational case. Notice that for  the subset of modes satisfying $m=n\theta$, the term proportional to $\epsilon_{mn}^*\alpha_{mn}$ in \eqref{SE-bo} vanishes. This makes the Hessian of $S_\mt{E}$ degenerate, implying the existence of zero modes. To properly account for the latter, notice that \eqref{SE-bo} splits as
\be\ba
S_\mt{E}&=S_0-\frac{ik}{(2\pi)^3}\sum_{m\neq n\theta}\left[(L_0n(m-n\theta)+\frac{i\beta}{4\pi}n^2(n^2+M_0))|\epsilon_{mn}|^2\right.\\
&\left.\quad
+(m-\theta n)(n^3+M_0n)\epsilon^*_{mn}\alpha_{mn}\right]+\frac{k\beta}{2(2\pi)^4}\sum_{n\theta\in\mathbb Z}n^2(n^2+M_0)|\epsilon_{n}|^2,
\ea
\ee
where $\epsilon_n\equiv\epsilon_{n\theta,n}$. The summation excludes $n=0,\,\pm 1$ for the vacuum state with $M_0=-1$ since its little group is ISO$(2,1)$, and excludes $n=0$ for states with $0>M_0>-1$, which is still compatible with rational $\theta$ (see \eqref{eq:theta}). Up to $2\pi$ factors, the one-loop partition function can be factorized into three parts 
\be\ba\label{1loop-special}
&\mathcal{Z}_{\mt{1-loop}}=\mathcal{Z}_{\mt{normal}}\mathcal{Z}_{\mt{special}}\int \prod_{n\theta\in\mathbb Z}d\alpha_{n},\quad \alpha_n\equiv\alpha_{n\theta,\theta},
\\&\mathcal{Z}_{\mt{normal}}=e^{-S_0}\prod_{m\neq n\theta}(m-n\theta)^{-1},\quad \mathcal{Z}_{\mt{special}}=\prod_{n\theta\in\mathbb Z}\left(\frac{(M_0+n^2)}{k\beta } \right)^{1/2}.
\ea
\ee
$\mathcal{Z}_{\mt{normal}}$ is the contribution from normal modes $m\neq n\theta$, so it has the same form as \eqref{Z-BMS} but with the product taken over $m\neq n\theta$. $\mathcal{Z}_{\mt{special}}$ counts the finite contribution from special modes with $m=n\theta$. The remaining factor $\int\prod_{m=n\theta}d\alpha_{mn}$ gives an  IR divergent factor.

\paragraph{Comments on one-loop exactness.} Before moving to the exact localization analysis, we would like to briefly comment on the approach and results recently reported on 1-loop exactness  in \cite{Cotler:2024cia}. In this work, it was noticed the linear functional dependence in $\tilde{\alpha}$ allows one to integrate it out exactly, leading to a delta functional of the $f$ mode. For irrational $\theta$, such localization leads to a unique saddle $f=f_0$, rendering the partition function 1-loop exact. Our result \eqref{Z-BMS} agrees with their conclusion and extends it to purely imaginary $\theta$. For rational $\theta$, there is a family of saddles $\{f_c\}$ satisfying the delta functional. As a result, the full partition function equals
\be
\mathcal{Z}=\int df\,\delta(\mathcal{F}[f])\,e^{-S_\mt{E}}=\sum_{f_c}\,e^{-S_\mt{E}[f_c]}\frac{1}{|\delta\mathcal{F}/\delta f|_{f_c}}\,,
\label{eq:cjetal}
\ee
with
\begin{equation}
    \mathcal{F}[f]=\partial_y\left(\{f,\varphi\}-\frac{M_0}{2}f'^2\right)\,.
\end{equation}
The result \eqref{1loop-special} should be recognized as the one-loop contribution at $f_c=f_0$. Computing \eqref{eq:cjetal} is difficult because it is hard to sum over all saddles $f_c$ as both $S_{\mt E}$ and the Jacobian $|\frac{\delta\mathcal{F}}{\delta f}|$ depend on $f_c$ in a complicated way. 

Lastly, one should not confuse the sum over saddles as discussed in \eqref{new-saddle} and the sum over $\{f_c\}$ as in \eqref{eq:cjetal}. The former is needed to compute the one-loop partition function, while the latter is needed to compute the \textit{full} partition function. By definition, the set $\{f_c\}$ does indeed solve the equation of motion obtained by varying $\tilde\alpha$. However, to claim these are indeed saddles, one still needs to show they exists a solution for $\tilde\alpha$ for the equations of motion obtained by varying $f$.

\subsection{Localization}

To examine the 1-loop exactness of the torus partition function,  we next explore the construction of Q-exact terms allowing us to localize the full path integral, as reviewed in section \ref{sec:review}. 

The first step is to identify the existence of some supersymmetry. As shown in appendix \ref{sec:bms-check}, after splitting $f=f_0 + \epsilon$, the full action $S_\mt{E}'=S_\mt{E} + S_\omega$
\be
\ba\label{SE-epsilon}
&S_\mt{E}=S_0-\frac{k}{2\pi}\int dyd\varphi \left[i(L_0\epsilon'+M_0\tilde\alpha')\p_y\epsilon-\frac{i\tilde\alpha'[(1+\epsilon')\p_y\epsilon''-\epsilon''\p_y\epsilon']}{(1+\epsilon')^3} -\frac{M_0}{2}\epsilon'^2-\frac{\epsilon''^2}{2(1+\epsilon')^2} \right] \\
 &S_\omega=\frac{k}{2\pi}\int dyd\varphi \left[L_0\psi'_f\psi_f+M_0\psi_\alpha'\psi_f-\frac{\psi_\alpha'(\psi_f''(1+\epsilon')-\psi_f'\epsilon'')}{(1+\epsilon')^3}+\frac{\tilde\alpha'\psi_f'\psi_f''}{(1+\epsilon')^3}\right]
\ea
\ee
is invariant under the supersymmetry transformations
\be\label{act-Q-bms3}
 Q\epsilon=\psi_f,\quad Q\tilde\alpha=\psi_\alpha,\quad Q\psi_f=-i\p_y\epsilon,\quad Q\psi_\alpha=\epsilon'-i\p_y\tilde\alpha\,.
\ee
Next, we discuss the construction of the localization term.

\subsubsection{Localization action} 
\label{sec:bms-laction}

As discussed around \eqref{eq:DH-cond}, we require Q-exact localization terms $QF$ that are Q-closed and have positive definite bosonic contributions. Consider the most general Grassmann-odd ansatz for the Q-exact localization term
\be\label{ansatz:QF-bms}
QF=
\int Q(\psi_f(D_1\epsilon+D_2\tilde\alpha)+\psi_\alpha(D_3\epsilon+D_4\tilde\alpha))
\ee
involving an arbitrary set of undetermined \textit{linear} operators $D_i$. Given the $Q^2$ action
\be
Q^2\epsilon=-i\p_y\epsilon,\quad Q^2\tilde\alpha=\epsilon'-i\p_y\tilde\alpha,\quad Q^2\psi_f=-i\p_y\psi_f,\quad Q^2\psi_\alpha=\psi_f'-i\p_y\psi_\alpha,
\ee
it follows
\be
\ba
Q^2F &=\int Q^2\psi_f(D_1\epsilon+D_2\tilde\alpha)+Q^2\psi_\alpha(D_3\epsilon+D_4\tilde\alpha)+\psi_fQ^2(D_1\epsilon+D_2\tilde\alpha)+\psi_\alpha Q^2(D_3\epsilon+D_4\tilde\alpha) \\
&=-\int i\p_y \left(\psi_f(D_1\epsilon+D_2\tilde\alpha)+\psi_\alpha(D_3\epsilon+D_4\tilde\alpha)\right) \\
& +\psi'_f\,D_3\epsilon +\psi_f\,D_2\epsilon' + \psi'_f\,D_4\tilde\alpha + \psi_\alpha\,D_4\epsilon'.
\ea
\ee
In order for $QF$ to be Q-closed, the above integrand must be a total derivative. The first line in the second equality is already of that form, whereas the conditions $D_2=D_3$ and $D_4=0$ achieve the same goal for the final line. The resulting Q-closed term can more explicitly be written as
\be
QF=-\int i\p_y\epsilon(D_1\epsilon+D_2\tilde\alpha)+(\epsilon'-i\p_y\tilde\alpha)D_2\epsilon+\psi_f(D_1\psi_f+D_2\psi_\alpha)+\psi_\alpha D_2\psi_f.
\ee
Letting\footnote{The factor $k$ is introduced to make all $a_i$ dimensionless.}
\be
D_1=k^{-1}a_1\p_y+ a_2\,\p_\varphi,\quad D_2=a_3\p_y+k a_4\,\p_\varphi,
\label{eq:lin-par}
\ee
and using \eqref{Fourier}, it follows 
\be
\ba\label{QF-BMS}
QF=&\sum_{m,n}-\frac{i(m-n\theta)(a_4nk\beta+2\pi a_3(m-n\theta))}{2\pi^2\beta}\epsilon_{mn}\alpha^*_{mn}
\\&+\frac{-4 i a_1 \pi^2 (m - n \theta)^2 + 
n k\beta \left(a_4 n k\beta - 2 i a_2 \pi (m - n \theta) + 
2 a_3 \pi (m - n \theta)\right)
}{4\pi^3k\beta}|\epsilon_{mn}|^2
\\&\frac{-i(a_2nk\beta+2\pi a_1(m-n\theta))}{4\pi^3k\beta}a_{mn}\wedge a^*_{mn}-\frac{a_4nk\beta+2\pi a_3(m-n\theta)}{4\pi^3\beta}a_{mn}\wedge b^*_{mn}.
\ea
\ee
The last step is to determine the coefficients $a_i$ in \eqref{eq:lin-par} to make the bosonic contribution to $QF$ positive definite. The latter can be written as
\be
\ba\label{QF-bosonic}
(QF)_\mt{bosonic}&=\sum_{m,n}\frac{A_{mn}}{2}(\epsilon_{mn}\alpha^*_{mn}+\epsilon^*_{mn}\alpha_{mn})+B_{mn}|\epsilon_{mn}|^2.
\ea
\ee
where $*$ stands for complex conjugate, as follows from the discussion below \eqref{eq:comp-conj}, and we defined the matrices
\be\ba
&A_{mn}=-\frac{i(m-n\theta)(a_4nk\beta+2\pi a_3(m-n\theta))}{2\pi^2\beta}\,, \\
&B_{mn}=\frac{-4 i a_1 \pi^2 (m - n \theta)^2 + 
n k\beta \left(a_4 n k\beta - 2 i a_2 \pi (m - n \theta) + 
2 a_3 \pi (m - n \theta)\right)
}{4\pi^3k\beta}\,.
\ea
\ee
In terms of the real degrees of freedom \eqref{eq:comp-conj}, \eqref{QF-bosonic} becomes
\be\ba\label{QF-real}
(QF)_\mt{bosonic}&=\sum_{m\geq0,n>0}{A_{mn}}(\epsilon^\mt{R}_{mn}\alpha^\mt{R}_{mn}+\epsilon^\mt{I}_{mn}\alpha^\mt{I}_{mn})+2B_{mn}((\epsilon^\mt{R}_{mn})^2+(\epsilon^\mt{I}_{mn})^2)
\\&=\sum_{m\geq0,n>0}E_{mn}M_{mn}E_{mn}^\mt{T}
\ea
\ee
where we assembled the different independent real modes into $E_{mn}=(\epsilon^R_{mn},\epsilon^I_{mn},\alpha^R_{mn},\alpha^I_{mn})$, allowing us to identify the matrix of Gaussian  fluctuations as
\be
M_{mn}=\left(\begin{array}{cccc}
 2 B_{mn}   & 0& A_{mn}&0 \\
  0   &2 B_{mn}&0&A_{mn}\\
  A_{mn}&0&0&0\\
  0&A_{mn}&0&0
\end{array}\right).
\ee
The matrix $M_{mn}$ has eigenvalues $(B_{mn}\pm\sqrt{A^2_{mn}+B^2_{mn}})$  with degenerate multiplicity 2. The convergence of the Gaussian integral requires the eigenvalues to have positive real parts. In our case, this is achieved by\footnote{Let $B=B^R+i B^I$ with $B^R>0,B^I\in\mathbb R$ and solve $\sqrt{A^2+B^2}=\tilde B^R+i\tilde B^I$ with $\tilde B^R,\tilde B^I\in\mathbb R$ and $A=i\tilde{A}$ being pure imaginary, we find $(\tilde B^R)^2=(\sqrt{4(B^I)^2(B^R)^2+((B^R)^2-(B^I)^2-\tilde{A}^2)^2}+((B^R)^2-(B^I)^2-\tilde{A}^2))/2>0$. It can be directly checked that $(\tilde B^R)^2-(B^R)^2<0$. As a result, for pure imaginary $A$ and $\text{Re}(B)>0$, $\text{Re}\,(B\pm\sqrt{A^2+B^2})>0$ does hold.}
\begin{equation}
  \text{Re}\,(B_{mn})>0 \qquad \text{and} \qquad A_{mn}\,\,\,\text{purely imaginary}
\label{eq:positivity}
\end{equation}
We shall distinguish two cases when solving these positivity requirements : real and purely imaginary $\theta$. 

\paragraph{Real $\theta$.} When $\theta$ is real, the positivity conditions \eqref{eq:positivity} can be achieved by 
\begin{equation}
  a_3,\,a_4 \in \mathbb{R}\,, \qquad a_4 \geq 0\,, \qquad a_1=i|a_1|\,, \qquad a_2 = -ia_3.
\end{equation}
Indeed
\be
B_{mn}=\frac{|a_1|}{\pi k\beta}\,(m-n\theta)^2 + a_4\,\frac{n^2k\beta}{4\pi^3}>0
\ee
is positive definite and $A_{mn}$ is purely imaginary, as required.

\paragraph{Purely imaginary $\theta$.} Requiring $A_{mn}$ to be purely imaginary is achieved by
\be
a_3=1,\quad a_4=\frac{4\pi \theta}{k\beta}\quad \Rightarrow \quad A_{mn}=-i\frac{(m^2+n^2|\theta^2|)}{\pi\beta}.
\ee
To analyse the positivity of $\text{Re}\left(B_{mn}\right)$, choose 
\begin{equation}
   a_1 = i|a_1| \qquad \text{and} \qquad  a_2 \in \mathbb{R}
\end{equation}
This leads to
\be
\text{Re}(B_{mn})=\frac{|a_1|}{\pi k\beta} \left(m+\frac{nk\beta}{4\pi |a_1|}\right)^2 - \frac{n^2}{4\pi^3k\beta} \left(\frac{k^2\beta^2}{4|a_1|} + 4\pi^2|a_1||\theta|^2 + a_2\,2\pi |\theta|\right)
\ee
whose positivity requires $a_2$ to satisfy
\be
  \frac{k^2\beta^2}{4|a_1|} + 4\pi^2|a_1||\theta|^2 + a_2\,2\pi |\theta|\ < 0\,.
\label{eq:img-pos}
\ee

\paragraph{Summary.}  A family of Q-exact $QF$ localization terms being Q-closed was explicitly constructed in \eqref{QF-BMS}. Among these, we showed the existence of subfamilies where by convenient choices of the constants determining the otherwise arbitrary linear operators, the positivity condition on the bosonic contribution to $QF$ was satisfied, both for $\theta$ real and purely imaginary. These are the relevant localization terms that we will use to perform the localization analysis in \ref{sec:l-analysis}.

\paragraph{Remark.} Before performing the full path integral using localization, let us briefly comment on an alternative way of computing some of our calculations following from our positivity conditions \eqref{eq:positivity}. When requiring the real parts of the $M_{mn}$ eigenvalues to be positive, one diagonalizes the matrix $M_{mn}$ to rewrite \eqref{QF-real} as  
\be
(QF)_{\mt{bosonic}}=\sum_{m\geq0,n>0}\tilde E_{mn}\Lambda_{mn}\tilde E^T_{mn}
\ee
with $\Lambda_{mn}$ diagonal. Since the new basis $\tilde E_{mn}$ is a complex linear combination of $E_{mn}$, one is deforming the contour of integration from $E_{mn}\in\mathbb R^4$ to $\tilde E_{mn}\in\mathbb R^4$, effectively leading to ordinary Gaussian integrals over real $\tilde E_{mn}$. 
However, when the conditions \eqref{eq:positivity} are satisfied, $A_{mn}$ is purely imaginary. It follows one could have performed the integrals over $\alpha_{mn}$ as a Fourier transformation, leading to
$|A_{mn}|^{-1}\delta(\epsilon_{mn})$, up to a proportionality constant. The latter localizes the mode $\epsilon_{mn}$ to 0. The two perspectives are equivalent and lead to same result. In the following discussion, we will adopt the first perspective with deformed contour.

\subsubsection{Localization analysis} 
\label{sec:l-analysis}

Once the fermionic localization term $QF$ is known, the next goal is to perform the path integral \eqref{Z-local}
\be\label{Z-local2}
\mathcal{Z}=\lim_{s\to\infty}\mathcal{Z}[s]=\int_{\mathcal M_{\mt{loc}}}[dx_c]\,e^{-S_{\mt E}[x_c]}\left.\frac{1}{\text{SDet}'(QF)}\right|_{x_c}\,,
\ee
where $x_c$ stands for the zeroes of the localization term $QF$, and then to
compare it with our 1-loop calculations in section \ref{sec:1-loop}. As in earlier discussions, our analyses distinguishes between irrational, or purely imaginary, $\theta$, and rational $\theta$.

\paragraph{Irrational or purely imaginary $\theta$.} The localization term \eqref{QF-BMS} is non-degenerate, leading to the unique zero
\be
  \mathcal{M}_{\mt{loc}}=\{\tilde E_{mn}=0,\forall m,n\}=\{\epsilon=\tilde\alpha=0\}.
\ee
The partition function \eqref{Z-local2} reduces to 
\be\label{Zs-irrational}
\mathcal{Z}=e^{-S_0}\left.\frac{1}{\text{SDet}'(QF)}\right|_{\mathcal{M}_{\mt{loc}}}.
\ee
Modulo $2\pi$ factors, the bosonic contribution to the superdeterminant equals 
\be
\ba
\int\left(\prod_{m,n}d\epsilon_{mn}d\alpha_{mn}\right) e^{-(QF)_\mt{bosonic}} & =\prod_{m\geq0,n>0}\det(M_{mn})^{-1/2} \\
&=\prod_{m\geq0,n>0}|A_{mn}|^{-2}=\prod_{m,n}|A_{mn}|^{-1}\,,
\ea
\ee
whereas the ghost fields contribution is
\be\label{Ghost}
\int\left(\prod_{m,n}da_{mn}db_{mn}\right) e^{-(QF)_\mt{ghost}}=\prod_{m,n}\frac{a_4nk\beta+2\pi a_3(m-n\theta)}{4\pi^3k\beta}\,.
\ee
Notice the dependence on the matrix $A_{mn}$ cancels, leading to the final result
\be\label{local-irrational-Z}
\mathcal{Z}=e^{-S_0}\prod_{m,n}(m-n\theta)^{-1}
\ee
This result matches the one-loop partition function \eqref{Z-BMS} computed perturbatively. This analysis proves that for irrational or purely imaginary $\theta$, the perturbative 1-loop calculation is exact. This is one of the main results in this paper.

\paragraph{Rational $\theta$.} The localization term  \eqref{QF-BMS} is degenerate when $\theta$ is rational. This gives rise to a non-trivial manifold of zero modes making the evaluation of the partition function \eqref{Z-local2} challenging. However, as we discuss next, the nature of the challenge depends on the choice of parameters labeling the family of localization terms  \eqref{QF-BMS}. Our next goal will be to identify a choice where the path integral can be performed exactly.

Consider the choice $a_4=0$. The bosonic contribution to the localization term \eqref{QF-bosonic} reduces to
\be
(QF)_{\mt{bosonic}}=\sum_{m\neq n\theta}A_{mn}\epsilon_{mn}\alpha^*_{mn}+ B_{mn}|\epsilon_{mn}^2|.
\ee
The set of critical points gives $\epsilon_{mn}=\alpha_{mn}=0$ \textit{only} for $m\neq n\theta$, while the modes $\epsilon_n\equiv\epsilon_{n\theta,n}$ and $\alpha_n\equiv\alpha_{n\theta,n}$ remain arbitrary. Similarly, the ghost contribution to \eqref{QF-BMS} is also degenerate, since for $m=n\theta$, the modes $a_n=a_{n\theta,n},b_n=b_{n\theta,n}$ remain arbitrary. Altogether, this leads to the localization submanifold 
\be
\mathcal{M}^{a_4=0}_{\mt{loc}}=\{(\epsilon(\varphi),\tilde\alpha(\varphi),a(\varphi), b(\varphi)\}
\label{eq:xc_a4}
\ee
in terms of four functions satisfying $h(\varphi+2\pi\theta)=h(\varphi+2\pi)=h(\varphi)$ for all $h$ choices. Plugging all this information into \eqref{Z-local2} and using the measure $[dx_c]=\prod_{n\theta\in\mathbb Z}d\epsilon_n d\alpha_n da_n db_n$, leads to
\be
\mathcal{Z}=\int \left(\prod_{m\neq n\theta}da_{mn}db_{mn}d\epsilon_{mn}d\alpha_{mn}\right) e^{-QF}\int_{\mathcal{M}^{a_4=0}_{\mt{loc}}} [dx_c]\,e^{-S_\mt E[x_c]-S_\omega}\,.
\ee
The evaluation of the partition function requires to integrate $S_\mt E[x_c]$ over $\mathcal{M}_{\mt{loc}}$. This is difficult since the value of both the euclidean action $S_{\mt E}[x_c]$ and the ghost action $S_\omega$ depend on $\epsilon$ non-linearly, as can be seen in \eqref{SE-epsilon}. This example illustrates the difficulty of performing the exact partition function within the localization technology.

As our second choice, let us explore $a_4\neq 0$. Setting $a_4=1$ for convenience, \eqref{QF-bosonic} equals
\be\label{QF-rational}
(QF)_{\mt{bosonic}}=\sum_{m\neq n\theta} \left(A_{mn}\epsilon_{mn}\alpha^*_{mn}+B_{mn}|\epsilon_{mn}|^2\right) +\sum_{n\theta\in\mathbb Z}\frac{n^2k\beta}{4\pi^3}|\epsilon_n|^2,
\ee
Notice the set of critical points $\{x_c\}$ involves $\epsilon_{mn}=0$ \textit{for all} $(m,n)$ and $\alpha_{mn}=0$ for $m\neq n\theta$. Using an analogous notation to the one introduced in \eqref{eq:xc_a4}, this set can be parametrized by
\be\label{xc}
\mathcal{M}^{a_4\neq 0}_{\mt{loc}}=\{(\epsilon=0,\tilde\alpha(\varphi))\} \qquad \text{with} \qquad \tilde\alpha(\varphi+2\pi\theta)=\tilde\alpha(\varphi+2\pi)=\tilde\alpha(\varphi).
\ee
Thus, the choice $a_4\neq 0$ localizes the phase space to a smaller submanifold compared to \eqref{eq:xc_a4}. A second advantage of the $a_4\neq 0$ choice is that $S_\mt{E}[x_c]=S_0$ \textit{for any} $\tilde\alpha(\varphi)$, allowing us to perform the integral over $[dx_c]$. Finally,
as long as $a_2\,a_4\neq 0$, the ghost part of $QF$ \eqref{QF-BMS} is non-degenerate, i.e. there are no further zero modes in this case. Altogether, the partition function \eqref{Z-local2} is given by
\be
\mathcal{Z}=\int\prod_{m,n}da_{mn}db_{mn} d\epsilon_{mn}\prod_{m\neq n\theta}d\alpha_{mn}e^{-QF}\int_{\mathcal{M}^{a_4\neq 0}_{\mt{loc}}}[dx_c]e^{-S_0}.
\ee
The contribution from the bosonic determinant, equals 
\be
\int\prod_{m,n}d\epsilon_{mn}\prod_{m\neq n\theta}d\alpha_{mn} e^{-(QF)_{\mt{bosonic}}}=\prod_{m\neq n\theta}|A_{mn}|^{-1}\prod_{n\theta\in\mathbb Z}(n^2k\beta)^{-1/2}\,,
\ee
where the last factor originates from integrating the second term in \eqref{QF-rational} over $\epsilon_n$. The integral over the ghost modes can also be split into normal modes $(m\neq n\theta)$
and $m=n\theta$ modes, leading to
\be
\int \prod_{m,n}da_{mn}db_{mn}e^{-(QF)_{\mt{ghost}}}=\prod_{m\neq n\theta}\frac{|A_{mn}|}{m-n\theta}\prod_{n\theta\in\mathbb Z}n
\ee
Altogether,
\be
\mathcal{Z}=e^{-S_0}\int\prod_{n\theta\in\mathbb Z}d\alpha_n\prod_{m\neq n\theta}\frac{1}{m-n\theta}\prod_{n\theta\in\mathbb Z}(k\beta)^{-1/2}.
\label{eq:final-Z}
\ee
Comparing with \eqref{1loop-special}, we find the exact partition function agrees with the 1-loop partition function evaluated around $f=f_0$ up to an overall factor which is independent of $\beta,\theta$,
\be
\mathcal{Z}=\mathcal{N}\mathcal{Z}_{\mt{1-loop}}
\ee
with
\be
\mathcal{N}=\prod_{n\theta\in\mathbb Z}\left(M_0+n^2\right)^{-1/2}
\ee

The partition function \eqref{eq:final-Z} is the last main result of this paper. It is remarkable how the use of the $a_4\neq 0$ localization term, compared to the $a_4=0$ one, allowed us to resum the contributions from the full set of saddle points given in \eqref{eq:cjetal}. Notice that stripping off its IR divergence, it would appear the remaining finite partition function would allow us to compute any relevant observables for 3d pure Einstein gravity.

\section{Summary of results}
\label{sec:summary}

The main result in this note is the computation of the torus partition function for the coadjoint orbit of  $\widehat{\text{BMS}}_3$ with constant representatives using the method of fermionic localization. As reviewed in section \ref{sec:review}, this requires to find a localization term $sQF$ in \eqref{eq:sQF} satisfying \eqref{eq:DH-cond}, i.e. being Q-closed and with positive definite bosonic contribution.  This term was formally constructed using the existence of a K$\Ddot{\text{a}}$hler metric on the symplectic spaces considered in \cite{Stanford:2017thb,Cotler:2018zff}. Since we were not aware of such structure for the coadjoint orbit of $\widehat{\text{BMS}}_3$, we constructed the relevant
localization term by making a proper ansatz and explicitly solving the conditions \eqref{eq:DH-cond}. 

This strategy was first tested in AdS$_3$. Starting with the ansatz \eqref{eq:Vir-loc-ansatz} and restricting to linear first order operators $D$, the conditions \eqref{eq:DH-cond} were explicitly solved leading to the  localization term  \eqref{QF-Vir-final}.  The resulting exact partition function matched the one-loop partition function \eqref{Z-Vir}, reproducing the well known 1-loop exactness result in the literature \cite{Cotler:2018zff}.

Next, the same strategy was applied for BMS$_3$. First, the symplectic form associated to the coadjoint orbit of $\widehat{\text{BMS}}_3$  with constant representative $(L_0,M_0)$ was computed in \eqref{eq:omega-bms3}. After writing the Pfaffian in terms of ghost fields, as in \eqref{eq:Zt1}, the supersymmetry of the full acton $S_\mt{E}'$ was given in  \eqref{act-Q-bms3}. To find a proper localization term $QF$, the ansatz \eqref{ansatz:QF-bms} was made. The condition $Q^2F=0$ is satisfied for $D_2=D_3,\,D_4=0$. This gives a family of localization terms parametrized by two arbitrary differential operators $D_1$ and $D_2$.  Restricting these to be linear differential operators, the positivity condition $(QF)_{\mt{bosonic}}\geq0$ was shown to be satisfied by imposing some conditions on the constant coefficients determining these linear operators (see section \ref{sec:bms-laction} for a more detailed discussion on these conditions).

Once the localization term was determined,  the exact torus partition function could, in principle, be performed. The computation depends on the value of the angular potential $\theta$. When $\theta$ is irrational or purely imaginary, the path integral localizes and the partition function equals \eqref{local-irrational-Z}. This matches the 1-loop partition function computed perturbatively in subsection \ref{sec:1-loop}, though it was already known in the literature \cite{Barnich:2015mui,Merbis:2019wgk}. This proves  the one-loop exactness of the BMS$_3$ torus partition function for  $\theta$ is irrational or purely imaginary.

However, when $\theta$ is rational, the path integral calculation is more subtle, both perturbatively and within the localization method. The subtleties are two fold. First, both the saddle points in the 1-loop calculation and the localization fixed points are not unique. They both span an infinite dimensional submanifold. Second, since both the original geometric action $S_{\mt E}$ and the localization term $QF$ are degenerate at quadratic order, these require careful treatment. 

These subtleties make the calculation of the complete perturbative one-loop partition function challenging. We reported the contribution around a very specific saddle $f_0$, the same one used in the irrational $\theta$ case, giving the result  \eqref{1loop-special}. When turning to the exact calculation using localization, we were able to bypass these difficulties by a specific choice of localization term \eqref{QF-BMS}, with $a_4\neq0$. This choice leads to the localization space given by \eqref{xc}. Since the resulting action turns out to be completely independent of $\tilde\alpha$, the integration over the localization space is trivial and leads to an IR divergent prefactor. Integrating over nonzero modes is tractable and  the final result is given by \eqref{eq:final-Z}. This agrees with the one-loop calculation at the single saddle \eqref{1loop-special} up to a factor independent of the torus modular parameters.

\paragraph{Acknowledgements.} We would like to thank Glenn Barnich and Nadav Drukker for useful discussions. JS was supported by the Science and Technology Facilities Council [grant numbers ST/T000600/1, ST/X000494/1].


\appendix

\section{Supersymmetry variations}
\label{sec:susy-check}

In this appendix, we present the details for checking the superymmetry invariance of the different actions considered in this work.

\subsection{Virasoro}
\label{sec:vir-check}

Once the Pfaffian of the symplectic form is implemented in terms of ghosts fields \eqref{eq:Zt1}, the geometric action $S_\mt{E}' = S_\mt{E} + S_\omega$ describing the coadjoint orbit of $\widehat{\text{Vir}}$ with constant representative $j_0$ is
\be
\ba
   S_\mt{E} &=\int dyd\varphi \left(j_0\,f'(f'+i\p_yf)+\frac{cf''(f''+i\p_yf')}{48\pi f'^2}\right)\,, \\
   S_\omega &=\int dyd\varphi \left( j_0\,\psi\psi'+\frac{c\psi'\psi''}{48\pi f'^2}\right)\,.
\ea
\ee
The latter is invariant under the supersymetry transformations
\be
Qf=\psi,\quad Q\psi=-f'-i\frac{\p f}{\p y}\,.
\ee
The proof is by explicit calculation
\be
\ba
QS'_\mt{E} &=\int j_0[\psi'(i\p_yf+f')+f'(i\p_y\psi+\psi')-\psi'(f'+i\p_yf)+\psi(f''+i\p_yf')] \\
 &+\frac{c}{48\pi}\left[\frac{\psi''(f''+i\p_yf')+f''(\psi''+i\p_y\psi')}{f'^2}-\frac{2f''(f''+i\p_yf')\psi'}{f'^3}\right] \\
 &+\frac{c}{48\pi}\left[\frac{\psi'(f'''+i\p_yf'')-\psi''(f''+i\p_yf')}{f'^2}\right] =\int (\p_\varphi+i\p_y)\left[j_0\psi f'+\frac{c}{48\pi}\frac{f''\psi'}{f'^2}\right]=0
\ea
\ee
and due to the variation of the original integrand being a total derivative.

As discussed around \eqref{AdS:act-exp}, it was more convenient for our fermionic localization analysis to split the zero mode from the function $f(y,\varphi)$, i.e. $f(y,\varphi) = f_0 + \epsilon(y,
\phi)$, and to work directly in terms of the periodic functions $\epsilon(y,\phi)$ and $\psi(y,\varphi)$. The full action becomes
\be
 S'_\mt{E}=S_0+\int dyd\varphi \left(j_0\epsilon'(i\p_y\epsilon+\epsilon')+\frac{c}{48\pi}\frac{\epsilon''(\epsilon''+i\p_y\epsilon')}{(1+\epsilon')^2}+j_0\psi\psi'+\frac{c}{48\pi}\frac{\psi'\psi''}{(1+\epsilon')^2}\right)
\ee
and its supersymmetry transformations are
\be
Q\epsilon=\psi,\quad Q\psi=-\epsilon'-i\p_y\epsilon\,.
\ee
Once more, this is shown by explicit computation
\be
\ba
QS_\mt{E}'&=\int j_0[\psi'(i\p_y\epsilon+\epsilon')+\epsilon'(i\p_y\psi+\psi')-\psi'(\epsilon'+i\p_y\epsilon)+\psi(\epsilon''+i\p_y\epsilon')] \\
 &+\frac{c}{48\pi}\left[\frac{\psi''(\epsilon''+i\p_y\epsilon')+\epsilon''(\psi''+i\p_y\psi')}{(1+\epsilon')^2}-\frac{2\epsilon''(\epsilon''+i\p_y\epsilon')\psi'}{(1+\epsilon')^3}\right] \\
 &+ \frac{c}{48\pi}\left[ \frac{\psi'(\epsilon'''+i\p_y\epsilon'')-\psi''(\epsilon''+i\p_y\epsilon')}{(1+\epsilon')^2}\right]
\\&=\int (\p_\varphi+i\p_y)\left[j_0\psi \epsilon'+\frac{c}{48\pi}\frac{\epsilon''\psi'}{(1+\epsilon')^2}\right]=0\,,
\ea
\ee
since the variation of the integrand remains a total derivative.

\subsection{BMS$_3$}
\label{sec:bms-check}

Proceeeding as in the Virasoro discussion, once the Pfaffian term in hte path integral is written in terms of ghost fields, see \eqref{eq:Zt1}, the full geometric action $S_\mt{E}' = S_\mt{E} + S_\omega$ equals
\be
\ba
S_\mt{E}  &=-\frac{k}{2\pi}\int dyd\varphi \left[i(L_0f'+M_0\tilde\alpha')\p_yf-\frac{i\tilde\alpha'(f'\p_yf''-f''\p_yf')}{f'^3}-\frac{M_0}{2}f'^2+\{f,\varphi\}\right] \\
S_\omega&=\frac{k}{2\pi}\int dyd\varphi \left[L_0\psi'_f\psi_f+M_0\psi_\alpha'\psi_f-\frac{\psi_\alpha'(\psi_f''f'-\psi_f'f'')}{f'^3}+\frac{\tilde\alpha'\psi_f'\psi_f''}{f'^3}\right]\,.
\ea
\ee
The latter is invariant under the supersymmetry transformations
\be
Qf=\psi_f,\quad Q\tilde\alpha=\psi_\alpha,\quad Q\psi_f=-i\p_y f, \quad Q\psi_\alpha=f'-i \p_y\tilde\alpha
\ee
The proof is by direct calculation. First,
\be\ba
QS_\mt{E}&=-\frac{k}{2\pi}\int dyd\varphi\,\left\{i\left[L_0(\psi_f'\p_yf+f'\p_y\psi_f)+M_0(\psi_\alpha'\p_yf+\tilde\alpha'\p_y\psi_f+if'\psi_f')\right]\right. \\
&\left.-i\left[\frac{\psi_\alpha'(f'\p_yf''-f''\p_yf')+\tilde\alpha'(\psi_f'\p_yf''+f'\p_y\psi''_f-\psi_f''\p_yf'-f''\p_y\psi_f')}{f'^3}\right]\right. \\
&\left.-i\,\frac{3\tilde\alpha'(f'\p_yf''-f''\p_yf')\psi_f'}{f'^4}
-\frac{f''\psi_f''}{f'^2}+\frac{f''^2\psi_f'}{f'^3}\right\}\,.
\ea\ee
Second,
\be\ba
QS_\omega&=\frac{k}{2\pi}\int dyd\varphi\, \left\{iL_0(-\p_yf'\psi_f+\psi_f'\p_yf)+M_0([f'-i\p_y\tilde\alpha]'\psi_f+i\psi_\alpha'\p_yf)\right. \\
& \left.+i\frac{\psi_\alpha'(f'\p_yf''-\p_yf'f'')}{f'^3}-\frac{f''(\psi_f''f'-\psi_f'f'')}{f'^3} \right.
\\
& \left.-i\frac{\p_y\tilde\alpha'(\psi_f''f'-\psi_f'f'')}{f'^3}-i\frac{\tilde\alpha'(\p_yf'\psi_f''-\psi_f'\p_yf'')}{f'^3}\right\}
\ea\ee
Summing both contributions, one finds
\be
QS_\mt{E}+QS_\omega=0\,.
\ee

Following the same philosophy as in the Virasoro analysis, it was convenient to split the zero mode $f_0$ in the main text, leading to the total geometric action $S_\mt{E}' = S_\mt{E} + S_\omega$ 
\be
\ba
&S_\mt{E}=S_0-\frac{k}{2\pi}\int dyd\varphi i\left[(L_0\epsilon'+M_0\tilde\alpha')\p_y\epsilon-\frac{\tilde\alpha'[(1+\epsilon')\p_y\epsilon''-\epsilon''\p_y\epsilon']}{(1+\epsilon')^3}\right]-\frac{M_0}{2}\epsilon'^2-\frac{\epsilon''^2}{2(1+\epsilon')^2}
\\&S_\omega=\frac{k}{2\pi}\int dyd\varphi L_0\psi'_f\psi_f+M_0\psi_\alpha'\psi_f-\frac{\psi_\alpha'(\psi_f''(1+\epsilon')-\psi_f'\epsilon'')}{(1+\epsilon')^3}+\frac{\tilde\alpha'\psi_f'\psi_f''}{(1+\epsilon')^3}
\ea
\ee
The latter is invariant under the supersymmetry transformations
\be
 Q\epsilon=\psi_f,\quad Q\tilde\alpha=\psi_\alpha,\quad Q\psi_f=-i\p_y\epsilon,\quad Q\psi_\alpha=\epsilon'-i\p_y\tilde\alpha
\ee
The proof is once more by direct calculation. First,
\be\ba
QS_\mt{E}&=-\frac{k}{2\pi}\int dyd\varphi\, \left\{i[L_0(\psi_f'\p_y\epsilon+\epsilon'\p_y\psi_f)+M_0(\psi_\alpha'\p_y\epsilon+\tilde\alpha'\p_y\psi_f+i\epsilon'\psi_f')]\right. \\
& \left.-i\left[\frac{\psi_\alpha'((1+\epsilon')\p_y\epsilon''-\epsilon''\p_y\epsilon')+\tilde\alpha'(\psi_f'\p_y\epsilon''+(1+\epsilon')\p_y\psi''_f-\psi_f''\p_y\epsilon'-f''\p_y\psi_f')}{(1+\epsilon')^3}\right.\right.\\
&\left.\left.-\frac{3\tilde\alpha'((1+\epsilon')\p_y\epsilon''-\epsilon''\p_y\epsilon')\psi_f'}{(1+\epsilon')^4}\right]-\frac{\epsilon''\psi_f''}{(1+\epsilon')^2}+\frac{\epsilon''^2\psi_f'}{(1+\epsilon')^3}\right\}\,.
\ea\ee
Second,
\be\ba
QS_\omega&=\frac{k}{2\pi}\int dyd\varphi\, \left\{iL_0(-\p_y\epsilon'\psi_f+\psi_f'\p_y\epsilon)+M_0([\epsilon'-i\p_y\tilde\alpha]'\psi_f+i\psi_\alpha'\p_y\epsilon)\right. \\
& \left.+i\frac{\psi_\alpha'((1+\epsilon')\p_y\epsilon''-\p_y\epsilon'\epsilon'')}{f'^3}-\frac{\epsilon''(\psi_f''(1+\epsilon')-\psi_f'\epsilon'')}{(1+\epsilon')^3}\right. \\
&\left. -i\frac{\p_y\tilde\alpha'(\psi_f''(1+\epsilon')-\psi_f'\epsilon'')}{(1+\epsilon')^3}-i\frac{\tilde\alpha'(\p_y\epsilon'\psi_f''-\psi_f'\p_y\epsilon'')}{(1+\epsilon')^3}\right\}
\ea\ee
Summing both contributions, one finds
\be
Q(S_\mt{E}+S_\omega)=0\,.
\ee


\bibliographystyle{fullsort}
\bibliography{sample}

\end{document}